\documentstyle{mn}

\newif\ifAMStwofonts

\ifoldfss
  \ifCUPmtlplainloaded \else
    \NewTextAlphabet{textbfit} {cmbxti10} {}
    \NewTextAlphabet{textbfss} {cmssbx10} {}
    \NewMathAlphabet{mathbfit} {cmbxti10} {} 
    \NewMathAlphabet{mathbfss} {cmssbx10} {} 
  \fi
  \ifAMStwofonts
    \ifCUPmtlplainloaded \else
      \NewSymbolFont{upmath} {eurm10}
      \NewSymbolFont{AMSa} {msam10}
      \NewMathSymbol{\upi}     {0}{upmath}{19}
      \NewMathSymbol{\umu}     {0}{upmath}{16}
      \NewMathSymbol{\upartial}{0}{upmath}{40}
      \NewMathSymbol{\leqslant}{3}{AMSa}{36}
      \NewMathSymbol{\geqslant}{3}{AMSa}{3E}

    \fi
  \fi
\fi 

\ifnfssone
  \newmathalphabet{\mathit}
  \addtoversion{normal}{\mathit}{cmr}{m}{it}
  \addtoversion{bold}{\mathit}{cmr}{bx}{it}
  \newmathalphabet{\mathbfit} 
  \addtoversion{normal}{\mathbfit}{cmr}{bx}{it}
  \addtoversion{bold}{\mathbfit}{cmr}{bx}{it}
  \newmathalphabet{\mathbfss} 
  \addtoversion{normal}{\mathbfss}{cmss}{bx}{n}
  \addtoversion{bold}{\mathbfss}{cmss}{bx}{n}
  \ifAMStwofonts
    \ifCUPmtlplainloaded \else
      \UseAMStwoboldmath
      \makeatletter
      \new@mathgroup\upmath@group
      \define@mathgroup\mv@normal\upmath@group{eur}{m}{n}
      \define@mathgroup\mv@bold\upmath@group{eur}{b}{n}
      \edef\UPM{\hexnumber\upmath@group}
      \new@mathgroup\amsa@group
      \define@mathgroup\mv@normal\amsa@group{msa}{m}{n}
      \define@mathgroup\mv@bold\amsa@group{msa}{m}{n}
      \edef\AMSa{\hexnumber\amsa@group}
      \makeatother
      \mathchardef\upi="0\UPM19
      \mathchardef\umu="0\UPM16
      \mathchardef\upartial="0\UPM40
      \mathchardef\leqslant="3\AMSa36
      \mathchardef\geqslant="3\AMSa3E
    \fi
  \fi
\fi 

\ifnfsstwo
  \DeclareMathAlphabet{\mathbfit}{OT1}{cmr}{bx}{it}
  \SetMathAlphabet\mathbfit{bold}{OT1}{cmr}{bx}{it}
  \DeclareMathAlphabet{\mathbfss}{OT1}{cmss}{bx}{n}
  \SetMathAlphabet\mathbfss{bold}{OT1}{cmss}{bx}{n}
  \ifAMStwofonts
    \ifCUPmtlplainloaded \else
      \DeclareSymbolFont{UPM}{U}{eur}{m}{n}
      \SetSymbolFont{UPM}{bold}{U}{eur}{b}{n}
      \DeclareSymbolFont{AMSa}{U}{msa}{m}{n}
      \DeclareMathSymbol{\upi}{0}{UPM}{"19}
      \DeclareMathSymbol{\umu}{0}{UPM}{"16}
      \DeclareMathSymbol{\upartial}{0}{UPM}{"40}
      \DeclareMathSymbol{\leqslant}{3}{AMSa}{"36}
      \DeclareMathSymbol{\geqslant}{3}{AMSa}{"3E}
    \fi
  \fi
\fi 

\ifCUPmtlplainloaded \else
  \ifAMStwofonts \else 
    \def\upi{\pi}
    \def\umu{\mu}
    \def\upartial{\partial}
  \fi
\fi

\title{ RR Lyrae variables in M5 as a test of pulsational theory}
\author[F. Caputo et al.]
       {F. Caputo$^1$, V. Castellani$^2$, M. Marconi$^1$ and V. Ripepi$^1$
\\
        $^1$ Osservatorio Astronomico di Capodimonte, 
Via Moiariello 16, 80131 Napoli, Italy \\ 
        $^2$ Dipartimento di Fisica, Piazza Torricelli 2, 56100, Pisa,
Italy}
\date{}

\pagerange{\pageref{firstpage}--\pageref{lastpage}}
\pubyear{1994}

\begin{document}

\maketitle

\label{firstpage}

\begin{abstract}
We present B and V CCD photometry for variables in the cluster central 
region, adding new data for 32 variables and giving 
suitable light curves, mean magnitudes and corrected colors for 17 RR
Lyrae.
Implementing  the data given in this paper with similar data already 
appeared in the literature we discuss a sample of 42 variables, 
as given by 22 RR$_{ab}$ and 20 RR$_c$, to
the light of recent predictions from pulsational theories. 
We find that the observational evidence concerning  M5 pulsators
appears in marginal disagreement with  predictions concerning the
color of the First Overtone Blue Edge (FOBE),
whereas a clear disagreement  appears between the ZAHB luminosities
predicted through evolutionary or pulsational theories.
\end{abstract}


\section{Introduction}

After about a century of investigation, the precise and detailed knowledge
of structural parameters of RR Lyrae variables in galactic globulars 
is still a relevant target for stellar astrophysics. As a matter of the
fact, 
the true luminosity of these variables is still under debate 
(see, e.g., Gratton et al. 1997, Caputo 1997, Cacciari 1999, 
and references therein), 
vis-a-vis the tantalizing evidence that  an improved knowledge of 
such an evolutionary parameter would produce firm
constraints about several relevant issues, as the distance
of the parent clusters and - in turn - on the distance to the
Magellanic
Clouds, which is a key--stone for assessing a reliable distance scale in
the Universe.
In this context, observational data for  these variables have become even
more   
interesting, since we are now facing  rather detailed scenarios of
theoretical 
evolutionary and pulsational predictions which obviously require 
to be tested to observations.

In a previous paper  (Brocato et al. 1996, hereinafter BCR96) 
we have already presented 
new light curves for 15 RR Lyrae stars in M5, pointing out 
the occurrence of several further pulsators 
in the crowded core of the cluster. In this paper we present data
for these variables in the cluster central region, adding 
new data for 32 variables and giving 
suitable light curves and mean magnitudes for 17 RR Lyrae.              
The next section  will present the explored fields, discussing
the existing literature on M5 variables to give suitable cross correlations
among the nomenclatures adopted by different authors and   
advancing a proposal for a homogeneous  cataloguing for these variables.
The new data will be discussed in section 3. The final section
will discuss  available data for M5 variables to
the light of recent predictions from pulsational theories.  
A brief conclusion will close the  paper.

\section{RR Lyrae in M5 central region}

Observational material was secured  in April 1989 with the 1.54 m. 
``Danish'' telescope at ESO (La Silla, Chile), equipped with a RCA CCD 
331 $\times$ 512 pixels. For details regarding observations, calibration 
and the reduction procedure the reader is referred to BCR96.
Fig.1 shows the four explored frames, giving the
identification map for the variables we will discuss in this paper. 
However, the nomenclature of the variables needs here 
to be discussed in some details.

\begin{figure}
 \vspace{3cm}
 \caption{Identification maps for the variables in the core of M5}
\end{figure}

As early as 1973, the Sawyer Hogg catalogue was already listing 
103 variables in M5, with 93 RR Lyrae. Since that time several 
further pulsators have been found, mainly in the crowded central region.
However,  the contemporary publication of several 
papers dedicated to these variables has produced some 
overlap in their identification and nomenclature, 
worth to be discussed and clarified. 

In BCR96 we presented 26  new cluster RR Lyrae variables, 
which were added to the Sawyer Hogg list, from V104 to V129. 
However, now we find that V124 was already marked as V102 in the
Sawyer Hogg list whereas 22 out of these 26 variables appear in the
list by Evstigneeva et al. (1995; [ESST]), who added 30 new variables
to the Sawyer Hogg catalogue (from V104 to V133) collecting
data from  Kadla et al. (1987: [KGYI]), Kravtsov, 
(1988, 1990, 1991, 1992: [K88], [K90], [K91], [K92]). In this series of
papers one can find positions 
and finding charts for the 30 new variables near the cluster core,
together with indications about periods and rather scattered 
light curves from photographic $B$ photometry. In the present investigation
we tested 4 out of the 8 stars in that list not in common with
BCR96.  As for the remaining four, V122, 126 and V127 
were not detected because of crowding, while V133 lies outside our fields. 
We confirm the variability for three out of the four objects, 
with the exception of V124 which does not show magnitude 
variations within 0.1 mag.
\par 

More recently, Reid (1996: [R96]) has published an extensive $V$, 
$I$ survey of RR Lyrae variables in M5, reporting light curves 
and mean magnitudes for 49 RR Lyrae, 44 from the Sawyer Hogg list and
5 which belong to the Evstigneeva extension. In addition, R96 lists
several RR Lyrae candidates in the cluster central region (Table 5 in R96).
We tested for variability all these candidates, 
detecting variability only in stars 315, which appears a RR Lyrae variable.
Moreover, both R96 and Yan \& Reid (1996: [YR96])  discovered several 
eclipsing binaries in the cluster field, all outside our explored region.

\par
In the meantime, Sandquist et al., (1996: [SBSH]) reported the 
discovery of 8 new variables which were added to the list of ESST 
with numbers from V134 to V141. No period was given. From our 
observations it appears that V134 was already named as V129 in ESST, 
V135 is a new RR, V137 is the star V118 in BCR96, and V139 is the star 315
in
R96 whereas we were not able to detect variability neither in V136 nor in 
V140 and V144. We can say nothing about V138 and V141 
because the former is saturated and the latter one lies out of our fields. 
Finally, on the basis of $F336W$ filter HST observations, 
Drissen \& Shara, (1998: DS) have recently given light curves for 
29 variables in the core of M5, with 5 new variables. 
Because of crowding, we were able to detect 
only one among those variables, which appears a RR Lyrae pulsator.  
\par
As a whole, one finds that the literature on M5 variables would greatly 
benefit by a reorganization of the above discussed observations. 
Table 1 summarizes the situation, giving in the first column our
adopted nomenclature which runs as follows:  from V104 to  V133 we 
follow the notation by  ESST, from V134 to V141 the classification by 
SBSH, arranging from V142 to V158 further variables from various authors.
Next columns give the cross identifications for all the variables in M5 
not in Sawyer Hogg (1973) catalogue, together with the source of the 
variable identifications. Last four columns give details for stars
in the field of the present investigation or, for objects outside our 
fields,  we give within brackets the original classification 
of the objects.

\par

\begin{table*}
\centering
\begin{minipage}{22cm}
\caption[]{Cross identification for  variable stars not included in the 
Sawyer Hogg  (1973) catalog. The column SOURCE gives the reference \\ 
for the first identification of the object. Last four columns give details
of present investigation, where X and Y are the coordinate in pixel \\ 
with respect to our reference frame}.
{\begin{tabular}{cccccccccc}
IDENT. & IDENT.   & IDENT. & IDENT       & IDENT  & SOURCE & X   & Y   &
FIELD & NOTES  \\
\noalign{\medskip}
THIS PAPER   & Paper II   & K88   & R96 or YR96 & DS   &         &pix  &
pix &       &   \\
\noalign{\bigskip}
V104 &  V114  &      & 333 & HST-V1  & KGYI     & 263.5 &  544.5&    F4 &
ecl.? \\
V105 &  V113  &  18  &     & HST-V10 & K88,KGYI & 256.9 &  486.4&    F3 &
RR Lyrae \\
V106 &  V112  &  17  &     & HST-V14 & K88,KGYI & 255.7 &  473.4&    F3 &
RR Lyrae\\
V107 &  V111  &  14  &     & HST-V24 & K88,KGYI & 245.3 &  444.4&    F3 &
RR Lyrae\\
V108 &  V110  &  13  &     & HST-V22 & K88,KGYI & 224.7 &  446.5&    F3 &
RR Lyrae\\
V109 &  V108  &   7  &     & HST-V20 & K88,KGYI & 202.7 &  454.7&    F3 &
RR Lyrae\\
V110 &  V107  &   6  &     &         & K88,KGYI & 192.5 &  443.7&    F3 &
RR Lyrae\\
V111 &  V121  &  10  &     &         & K88,KGYI & 192.3 &  486.2&    F3 &
RR Lyrae\\
V112 &  V106  &   3  &     &         & K88,KGYI & 185.1 &  382.6&    F3 &
RR Lyrae\\
V113 &  V105  &   4  &     &         & K88,KGYI & 184.7 &  378.0&    F3 &
RR Lyrae\\
V114 &  V104  &   5  &     &         & K88,KGYI & 180.7 &  445.2&    F3 &
RR Lyrae\\
V115 &  V117  &   1  &     &         & K88      & 145.6 &  460.9&    F3 &
RR Lyrae\\
V116 &  V119  &   2  & 229 &         & K88      & 145.6 &  444.5&    F3 &
RR Lyrae\\
V117 &  V120  &   8  &     &         & K88      & 191.7 &  457.8&    F3 &
RR Lyrae\\
V118 &  V122  &   9  &     & HST-V18 & K88      & 201.7 &  468.2&    F3 &
RR Lyrae\\
V119 &  V109  &   11 &     & HST-V11 & K88      & 216.8 &  485.2&    F3 &
RR Lyrae \\
V120 &  V125  &   12 &     & HST-V17 & K88      & 226.6 &  469.3&    F3 &
RR Lyrae\\
V121 &  V127  &   15 &     & HST-V26 & K88      & 255.1 &  437.7&    F3 &
RR Lyrae\\
V122 &        &   16 &     & HST-V23 & K88      &       &       &    F3 &
crowded\\
V123 &  V126  &   19 &     & HST-V5  & K88      & 244.7 & 515.0 &    F4 &
RR Lyrae\\
V124 &        &   20 &     &         & K88      & 265.8 &  518.0&    F4 &
constant?\\
V125 &  V128  &   21 &     & HST-V9  & K88      & 280.0 &  489.4&    F3 &
RR Lyrae\\
V126 &        &   22 &     & HST-V12 & K88      &       &       &    F3 &
crowded\\
V127 &        &   23 &     & HST-V13 & K88      &       &       &    F3 &
crowded\\
V128 &  V129  &   24 &     & HST-V25 & K88      & 323.9 &  437.0&    F3 &
RR Lyrae\\
V129 &        &   25 &     &         & K88      & 348.3 &  348.9&    F2 &
RR Lyrae\\
V130 &  V115  &   26 & 365 &         & K88      & 72.7  &  570.7&    F4 &
RR Lyrae\\
V131 &        &   27 & 369 &         & K88      & 72.3  &  565.4&    F4 &
RR Lyrae\\
V132 &        &      &     &         & K91      & 147.3 &  547.2&    F4 &
RR Lyrae \\
V133 &        &      & 612 &         & K91      &       &       &       &
(RR Lyrae)\\
V134 &        &      &     &         & SBSH     & 83.8  &  348.9&    F2 & =
V129 \\
V135 &        &      &     &         & SBSH     & 265.9 &  341.1&    F3 &
RR Lyrae  \\
V136 &        &      &     &         & SBSH     & 197.2 &  273.9&    F3 &
constant?\\
V137 & V118   &      &     &         & SBSH     & 145.5 &  536.5&    F4 &
RR Lyrae\\
V138 &        &      &     &         & SBSH     & 66.4  &  114.4&    F1 &
saturated\\
V139 &        &      & 315 & HST-V3  & SBSH     & 281   &  64.4 &    F4 &
RR Lyrae\\
V140 &        &      &     &         & SBSH     & 150.8 &  504.6&    F3 &
constant?\\
V141 &        &      &     &         & SBSH     &       &       &       &  
type? \\
V142 & V123   &      &     & HST-V21 & BCR96  & 205.1 &  453.6&    F3 & RR
Lyrae\\
V143 & V116   &      &     &         & BCR96  & 138.8 &  462.4&    F3 &
type?  \\
V144 &        &      & 596 &         & R96      & 287.9 &  551.8&    F4 &
constant?  \\
V145 &	      &	     & 629 &         & R96	&	&	&	& (ecl.)  \\
V146 &        &      & 648 &         & R96	&	&	&	& (ecl.) \\
V147 &        &      & 651 &         & R96	&	&	&	& (ecl.) \\
V148 &        &      & 652 &         & R96	&	&	&	& (ecl.)\\
V149 &        &      & V1  &         & YR96	&	&	&	& (ecl.) \\
V150 &        &      & V2  &         & YR96	&	&	&	& (ecl.) \\
V151 &        &      & V3  &         & YR96	&	&	&	& (ecl.)\\
V152 &        &      & V4  &         & YR96	&	&	&	& (ecl.) \\
V153 & 	      &	     & V5  &         & YR96	&	&	&	& (ecl.)\\
V154 &	      &	     & V6  &         & YR96	&	&	&	& (ecl.)\\
V155 &	      &	     &     & HST-V6  & DS	& 	& 	&   F4	& crowded \\
V156 &	      &	     &     & HST-V15 & DS	& 	& 	&   F3	& crowded\\
V157 &        &	     &     & HST-V16 & DS	& 	&	&   F3	& crowded \\
V158 &        &	     &     & HST-V19 & DS	& 258.6	& 454.6	&   F3	& RR
Lyrae\\
V159 &        &	     &     & HST-V28 & DS	&	&	&   F3	& crowded\\
\end{tabular}}
\end{minipage}
\end{table*}

According to the data summarized in Table 1, one finds that out of 44
objects nominally present in our field 1, V134 is a duplicated
nomenclature, in four cases (V124, V136, V140 and V144) we failed 
to detect luminosity variations, 
7 objects are in a too crowded region and, finally, the image of a further 
object appears saturated. This paper will deal with the data for the 
remaining 31 variables plus V103 which falls in the investigated fields.
The first column in Table 2 gives the list of these  objects.

\begin{table*}
\centering
\begin{minipage}{15cm}
\caption[]{Periods, Epochs and amplitudes for all the new variables
investigated. 
See text for details on individual stars.}
\begin{tabular}{lllllllc}
VAR & Type&  Period & Epoch & A(V) & A(B) & Period Source & Notes   \\
\noalign{\bigskip}
V103 &  RRab &  0.5667   & 47629.551  & 0.77  & 1.03    & K92        & \\
V104 & 	ecl? &	0.741?   &            &	$>0.7$& $>0.9$  & THIS PAPER & \\	
V105 &	RRc  &	0.2920   & 47629.813  &	0.70  &	0.94	& THIS PAPER & \\		
V106 &	RRab &	0.5624   & 47629.311  &	0.95  &	1.20	& THIS PAPER & \\		
V107 &	RRab &	0.5117   & 47629.776  &	1.18  &	1.52	& K90        & \\		
V108 &	RRc  &  0.329    &            &	0.40  &	0.55	& THIS PAPER & \\		
V109 &	RRab &	0.476    & 47629.854  &	1.2   &	1.55	& THIS PAPER & \\
V110 &	RRab &	0.599    &            &	0.75  &	0.95	& K90        & \\	
V111 &	RRab &	0.6233   & 47629.283  &	0.63  &	0.93	& THIS PAPER & \\
V112 &	RRab &	0.5367   & 47629.546  &	0.97  &	1.29	& THIS PAPER & \\
V113 &	RRc  &	0.2843   & 47629.758  &	0.58  &	0.70	& THIS PAPER & \\
V114 &	RRab &	0.6037   & 47629.573  &	0.84  &	1.08 	& K90        & \\
V115 &	RRab &	0.6034   & 47629.772  &	0.57  &	0.77	& THIS PAPER & \\
V116 &	RRc  &	0.347288 & 47629.518  &	0.50  &	0.59	& R96        & \\
V117 &	RRc  &	0.3350   & 47629.520  &	0.4   &	0.5	& K90        & \\
V118 &	RRab &	0.5805   & 47629.832  &	1.09  &	1.57	& THIS PAPER & \\
V119 &	RRab &	0.5629   & 47629.297  & 0.95  &	1.25	& THIS PAPER & \\
V120 &	RRc  &	0.2797   & 47629.698  &	0.57  &	0.71	& THIS PAPER & \\
V121 &	RRab &	0.5865   & 47629.303  &	1.33  &	1.69	& THIS PAPER & \\
V123 &	RRab?&	0.6025   &            &$>0.5$ &$>0.65$	& K90        & \\
V125 &	RRc  &	0.3065   & 47629.592  &	0.49  &	0.65	& THIS PAPER & \\
V128 &	RRc  &	0.3080   & 47629.785  &	0.54  &	0.68	& THIS PAPER & \\
V129 &  RRab &	0.6011   & 47629.631  &	0.57  &	0.74  	& THIS PAPER & \\
V130 &	RRc  &	0.327396 & 47581.469  &	0.4   &	0.5	& R96        & \\
V131 &	RRc  &	0.281521 & 47629.763  &	0.54  &	0.68	& R96        & \\
V132 &	RRc  &	0.2835   & 47629.648  &	0.39  &	0.49	& THIS PAPER & \\
V135 &	RRab &	0.6304   & 47629.556  &	0.57  &	0.75	& THIS PAPER & \\
V137 &  RRab?&	0.591    & 47629.520  &	0.5   &	0.6	& THIS PAPER &
alternative period p=0.374\\
V139 &	RRc  &	0.300    & 47629.587  &	0.34  &	0.42	& THIS PAPER & \\
V142 &	RRab &	0.4577   & 47629.830  &	1.0   &	1.5	& THIS PAPER & \\
V143 &	     &	         &	      &	$>0.3$&	$>0.4$  & &  unknown variable type
\\
V158 &  RRc? &  0.45     &            &       &         & DS & \\
\end{tabular}
\end{minipage}
\end{table*}

\subsection{Light curves and notes on individual variables}

In order to obtain light curves for the  observed  variables, we searched
for periods by using a standard discrete Fourier 
decomposition (Deeming 1975, Kurtz 1985). However, in some  
cases the restricted time allowed for the observations
caused the occurrence of too few phase points along the pulsational cycle. 
In that case we adopted  periods from the literature. 
Data on Periods, Epochs of maximum, 
$B$ and $V$ amplitudes are reported in the same Table 2, together 
with the period source for each variable. Light curves for all 
variables listed in Table 2 are shown in Figure 2. HJD and $BV$ magnitudes
will be available via anonymous ftp at the address: oacosf.na.astro.it
(pub/M5RR) or upon request by e-mail. 

\begin{figure}
\vspace{3cm}
\caption{Light curves for all variable stars in our fields.}
\end{figure}
 
>From the sample of RR Lyrae in Table 2 we selected 17 objects with the
best 
light curves allowing a suitable estimate of mean magnitudes and colors. 
For this purpose, data for the selected objects have been  best-fitted
with a spline function and then integrated over the fitted curve 
to find mean quantities by averaging i) over the magnitude 
curve and ii) over the intensity curve. The results of this 
procedure are reported in Table 3, 
where individual columns give: (1) variable name; (2) period in days; 
(3) magnitude and (4) intensity averaged $V$; 
(5) magnitude and (6) intensity averaged $B-V$; 
(7)  corrected $(B-V)_c$ colors 
following Bono et al.\ (1995: [BCS]) prescriptions; (8) visual and 
(9) blue amplitudes.

As discussed by BCS, $<V>$ appears a good indicator for the magnitude of 
the ``static star'', whereas the observed averaged colors need to 
be corrected in order to give the ``static'' color.   
The consistency of BCS corrections is displayed in Fig. 3, where the 
upper panel shows the comparison between observed $<B> - <V>$ and $(B-V)$ 
colors and the lower panel gives the same comparison but after 
applying BCS corrections to the observed data. The agreement between 
the two corrected colors is almost perfect.

\begin{figure}
\vspace{3cm}
\caption{Upper panel: The comparison between raw $<B>-<V>$ and $(B-V)$ 
colors. Lower panel: The same, but for amplitude corrected data, according
to BCS.}
\end{figure}

\begin{table*}
\centering
\begin{minipage}{10cm}
\caption[]{Periods, mean visual magnitudes, mean colors and amplitudes 
for all the new variables investigated.}  
\begin{tabular}{ccccccccc}
 STAR & Period & $(V)$ & $<V>$ & $(B-V)$ & $ <B> - <V> $ & $ (B-V)_{c}$ &
A(V) & A(B) \\
\noalign{\medskip}
 (1)  &  (2)   & (3) & (4) & (5) & (6) & (7) & (8) & (9) \\
\noalign{\bigskip} 
V103 & 0.5667  & 15.132 & 15.109 &  0.433 & 0.410 & 0.423 & 0.77 & 1.03 \\
V105 & 0.2920  & 15.369 & 15.343 &  0.228 & 0.210 & 0.208 & 0.70 & 0.94 \\
V111 & 0.6233  & 15.007 & 14.986 &  0.438 & 0.412 & 0.428 & 0.63 & 0.93 \\
V112 & 0.5367  & 15.106 & 15.064 &  0.339 & 0.306 & 0.329 & 0.96 & 1.26  \\
V113 & 0.2843  & 15.173 & 15.154 &  0.152 & 0.142 & 0.143 & 0.58 & 0.70 \\
V114 & 0.6037  & 15.218 & 15.190 &  0.394 & 0.373 & 0.384 & 0.84 & 1.08 \\
V115 & 0.6034  & 14.999 & 14.985 &  0.466 & 0.452 & 0.456 & 0.57 & 0.77 \\
V116 & 0.34729 & 14.925 & 14.912 &  0.293 & 0.286 & 0.288 & 0.50 & 0.59 \\
V120 & 0.2797  & 15.179 & 15.160 &  0.260 & 0.250 & 0.251 & 0.57 & 0.71 \\
V121 & 0.5865  & 15.402 & 15.328 &  0.418 & 0.362 & 0.399 & 1.33  & 1.69 
\\
V125 & 0.3065  & 15.133 & 15.120 &  0.317 & 0.306 & 0.310 & 0.49 & 0.65 \\
V128 & 0.3080  & 15.065 & 15.048 &  0.293 & 0.283 & 0.285 & 0.54 & 0.68 \\
V129 & 0.6011  & 15.136 & 15.121 &  0.450 & 0.438 & 0.441 & 0.57 & 0.74 \\
V131 & 0.28152 & 15.069 & 15.052 &  0.248 & 0.237 & 0.240 & 0.54 & 0.68 \\ 
V132 & 0.2835  & 14.972 & 14.963 &  0.247 & 0.241 & 0.247 & 0.39 & 0.49 \\
V135 & 0.6304  & 15.083 & 15.068 &  0.447 & 0.432 & 0.437 & 0.57 & 0.75 \\
V139 & 0.300   & 14.861 & 14.854 &  0.115 & 0.112 & 0.112 & 0.34 & 0.42 \\
\noalign{\smallskip} 
\end{tabular}
\end{minipage}
\end{table*}

\noindent
Comments on individual stars, are as follows:
\begin{itemize}

\item
{\bf  V104}: this relevant object has been classified by R96 as a 
probable double mode RR Lyrae with a fundamental period P=0$^d$.464243.
However, 
DS reports this variable (their HST-V1) as contact 
or semi-detached binary, with a period larger than $0^d$.47. 
Our observations agree with DS, giving 
a period of about 0.74 days. The light curve by R96 shows 
both an eclipsing binary-like and a RR Lyrae-like shape, and our conclusion
is
that the nature of V104 remains a mystery, deserving more observations. 

\item
{\bf  V105}: the star shows a visual magnitude significantly fainter 
($\sim$0.3 mag) than other RR mean magnitudes, with a color 
($B-V\sim0.20$) in agreement with the blue edge of the instability strip. 
Possibly a foreground variable.
  
\item
{\bf V113}: this star shows a color which is 
significantly bluer that the blue edge for first--overtone pulsation 
but a visual magnitude in agreement with the average luminosity 
of the other variables. This evidence is not consistent with an 
explanation in terms of blending effect, nor  
this star is embedded in a crowded field (see Fig. 1). Accordingly, 
the color of V113 remains unexplained.

\item
{\bf V121}: this star shows a visual magnitude significantly fainter
($\sim$0.3 mag) than the mean magnitude but a rather red color. 

\item
{\bf V139}: again a peculiar variable; the color is bluer than the 
blue edge of the instability strip and its visual magnitude is about 0.2
mag 
brighter than the mean value. In this case a blending effect (due to a very
blue HB companion) could be invoked. 

\item
{\bf V143}: from the HJD vs. magnitude plot (see Fig. 2) 
this star appears clearly variable even though 
we were unable to find any periodicity. Thus also the nature of this 
star remains uncertain.

\end{itemize}

Our complete sample (this paper + BCR96) contains 11 RR Lyrae 
in common with R96, namely, in R96 notation, V36, V45, V81, V82, V97
 ($ab$--type variables) and V35, V57, V78, V100, V116, V131 
($c$--type variables). Figure 4 compares 
$<V>$ and $A(V)$ for these common objects. 
>From the upper panel of the figure one finds that our observations 
appear on the average fainter in $<V>$ by only $\sim$ 0.03 mag. As for  
visual amplitudes (lower panel), we have similar results with the
remarkable exception of V97. However, a simple inspection of R96 
light curve for V97 reveals that this star shows a Blazhko effect, 
as many others in this sample.

\begin{figure}
\vspace{3cm}
\caption{Comparison between present and R96 photometry.}
\end{figure}

Implementing  the data given in this paper with similar data already 
presented in BCR96 and with the $B$,$V$ photometry given by
Storm et al. (1991: [SCB]) for 11 RR Lyrae in the cluster periphery, 
we are eventually  dealing with  reliable light curves for  
a sample of 42 variables, as given by 22 RR$_{ab}$ 
and 20 RR$_c$. This sample, with the large occurrence
of core RR Lyrae variables, appears abnormally rich of c-type pulsators,
with the ratio $N_{c}$/$N_{RR}\sim$ 0.5 against the value of
0.26 found from the Sawyer Hogg catalogue (see, e.g., Castellani \& Quarta
1987).
However, this is due, at least in part,
to the marked asymmetry of $ab$ light curves which not allowed
a suitable coverage of the pulsational cycle within the too restricted sets
of observational data, nor the rather small number of objects gives 
a strong statistical significance to this ratio.

It is obviously interesting to compare
the pulsational behavior of our sample with the behavior of
variables in the outer regions of the cluster. Moreover, the availability
of magnitudes and colors of the "static" structures will allow to 
investigate  the color-magnitude (CM) location of the variables.
As for the first point, Figure 5 shows the Bailey diagram
(visual amplitude $A(V)$ $vs$ period)
for our sample, together with previous data for regularly
pulsating variables (no Blazhko effect) from SCB and
R96 studies. 
The remarkable similar distribution of stars in the various samples
shows that RR Lyrae in the various cluster regions have a 
common pulsational behavior.

\par
The CM diagram of our complete sample of 
22 RR$_{ab}$ and 20 RR$_c$ variables is shown in Fig. 6 together 
with the data for static HB stars by Brocato et al. (1995). 
By excluding V113 and V139 for the peculiar blue color, 
and V105 and V121 for the peculiar faint magnitude, one derives that  
the edges of the observed instability strip occur
at $(B-V)$=0.20$\pm$0.02 and 0.45$\pm$0.02 , as well as 
that the lower envelope of the RR Lyrae distribution can be safely put at 
$V$=15.14$\pm$0.04 mag. 
Note that we are well confident in the colors of the RR Lyrae near the
edges of the instability strip, whereas the few non-variable stars within
the 
instability strip appear in general affected by crowding or blending. 
\par

\begin{figure}
\vspace{3cm}
\caption{Bailey amplitude-period diagram for the labeled samples of 
M5 variables, in comparison with the results of convective pulsating 
models.}
\end{figure}

\begin{figure}
\vspace{3cm}
\caption{CM diagram of RR Lyrae and HB static stars in
M5; data for static stars are from Brocato et al. (1995); solid lines 
represent the reasonable  range for the ZAHB luminosity level.}
\end{figure}

\section{Discussion and conclusions}

In the previous figure 5 observational data are compared with 
theoretical predictions from
pulsating convective models with mass $M=0.65M_{\odot}$, luminosity
log$L/L_{\odot}$=1.61 (solid line) and 1.72 (dashed line) and the
labeled chemical composition. One finds that the left envelope of 
the RR$_{ab}$ 
distribution appears in agreement with the log$L/L_{\odot}$=1.61 line. 
Taking the pulsational results to their face value, this would suggest a 
ZAHB luminosity in excellent agreement with theoretical predictions given
by Castellani et al. (1991) but somewhat lower than the value 
log$L/L_{\odot}$=1.66 suggested by the more recent "improved" computations 
by Cassisi et al. (1998: [C98]). Note that all the most recent evaluations
of such a luminosity, as given by various authors under slightly different
assumptions, appear in the range 1.66$\pm$0.01 
(Caloi et al. 1997, Straniero et al. 1997). 

To discuss such a disagreement, one has to notice that while the
luminosity of the HB appears as a rather firm theoretical 
prediction, theoretical pulsator masses
rely on the assumption of a solar-like distribution of heavy elements. 
However, the suggested occurrence
of $\alpha$ enhanced mixtures in metal poor globular
cluster stars (see, e.g., Gratton et al. 1997) is not of help, 
since lower pulsator masses  
will imply even lower "pulsational" luminosities 
(see, e.g., figure 15 in Bono et al. 1997a).

In order to compare theoretical predictions about the boundaries
of the instability region with the observational results given
in the previous section one needs an estimate of the cluster 
distance modulus. Let us here assume the value one would
derive from the  already quoted C98 evolutionary results, namely
$DM$=14.59 ($\pm$0.04) mag. On this ground, Figure 7 compares 
the CM diagram location of the variables with 
theoretical predictions evaluated for the purpose of this paper 
(with the same code and input physics as in Bono et al., 1997b) 
about the boundaries of the instability strip for 
$M=0.65 M_{\odot}$, $Z$=0.001 (solar scaled), $Y$=0.24 
and for selected assumptions about the cluster reddening. 
The theoretical boundaries have been transformed into the 
observational plane by adopting the Castelli et al. (1997a,b) 
static model atmospheres.

One finds that the the fitting between theory and observation
would require a negligible amount of reddening,
running against a rather well established reddening of the order of
E(B-V)$\sim$0.02 (see, e.g., R96), if not larger (Gratton et al. 1997). 
Quite similar results are obtained 
when R96 data are compared with pulsational predictions in $(V-I)$ colors. 
Reasonable variations in the distance modulus do not play a
relevant role, because of the mild dependence
of the blue edge color on luminosity, nor the  occurrence
of $\alpha$ enhanced mixtures will be again of help, since lower pulsator
masses will imply even cooler blue boundaries (see again Bono et al. 1997a).

\begin{figure}
\vspace{3cm}
\caption{CM diagram of RR Lyrae stars compared with 
the theoretical boundaries of the instability, for 
different assumptions about the cluster reddening.}
\end{figure}

However, one finds that the problem arises from the adoption
of new model atmospheres by Castelli et al. (1997a,b),
since model atmospheres by Kurucz (1992) adopted in previous works
like, e.g., in Bono et al. (1997) would give for M5 a "pulsational"
reddening of the order of E(B-V)$\sim$0.02 mag. Since it is not clear to us
if "new model atmospheres" means in all cases "better models", and 
taking also into account reasonable ($\pm0.02$ mag) errors in 
the pulsator colors, we 
conclude for a suggestion (rather than an evidence)  for too
cool FOBE predictions from the adopted theoretical pulsational scenario.

As a conclusion, observational evidence concerning the M5 pulsators
appears in marginal disagreement with the observed color of FOBE,
whereas a clearer disagreement appears between the ZAHB luminosities
predicted through evolutionary or the theoretical Bailey diagram. 
Before closing the paper, we will notice that one should  be not 
too surprised about the quoted disagreements: in all cases we are 
dealing with rather small quantities ($\Delta M_V$ not larger than  0.1 mag, 
$\Delta(B-V) \sim 0.02$ mag) and, in that sense, both 
evolutionary and pulsational theories appear in a rather satisfactory 
agreement with observations.
Conversely, when dealing -as we do -with fine details of theoretical
predictions, one should be surprised if these results do not show
any evidence of the many uncertainties we know are existing
in the theoretical scenario.
\bigskip

\noindent
{\bf Acknowledgment:} 
We wish to thank Dr. Kravtsov who sent us that part of his 
{\it Candidate's dissertation} regarding M5 variables, and 
E. Brocato
and A. Weiss for
useful discussion. 
This work was partially supported by Consorzio Nazionale per 
l'Astronomia e l'Astrofisica (C.N.A.A.) through a postdoc research 
grant to M. Marconi.

\bsp

\label{lastpage}

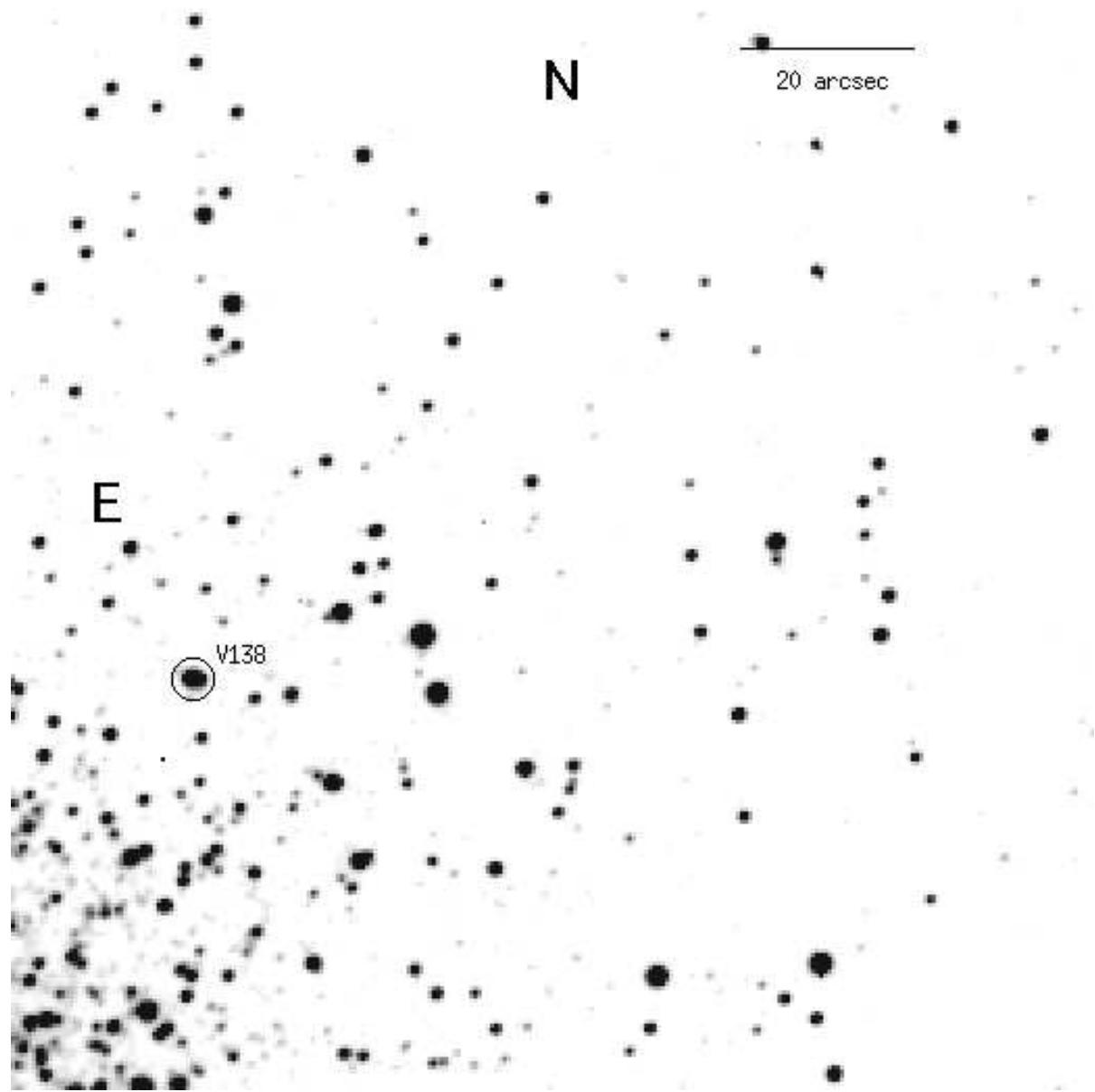
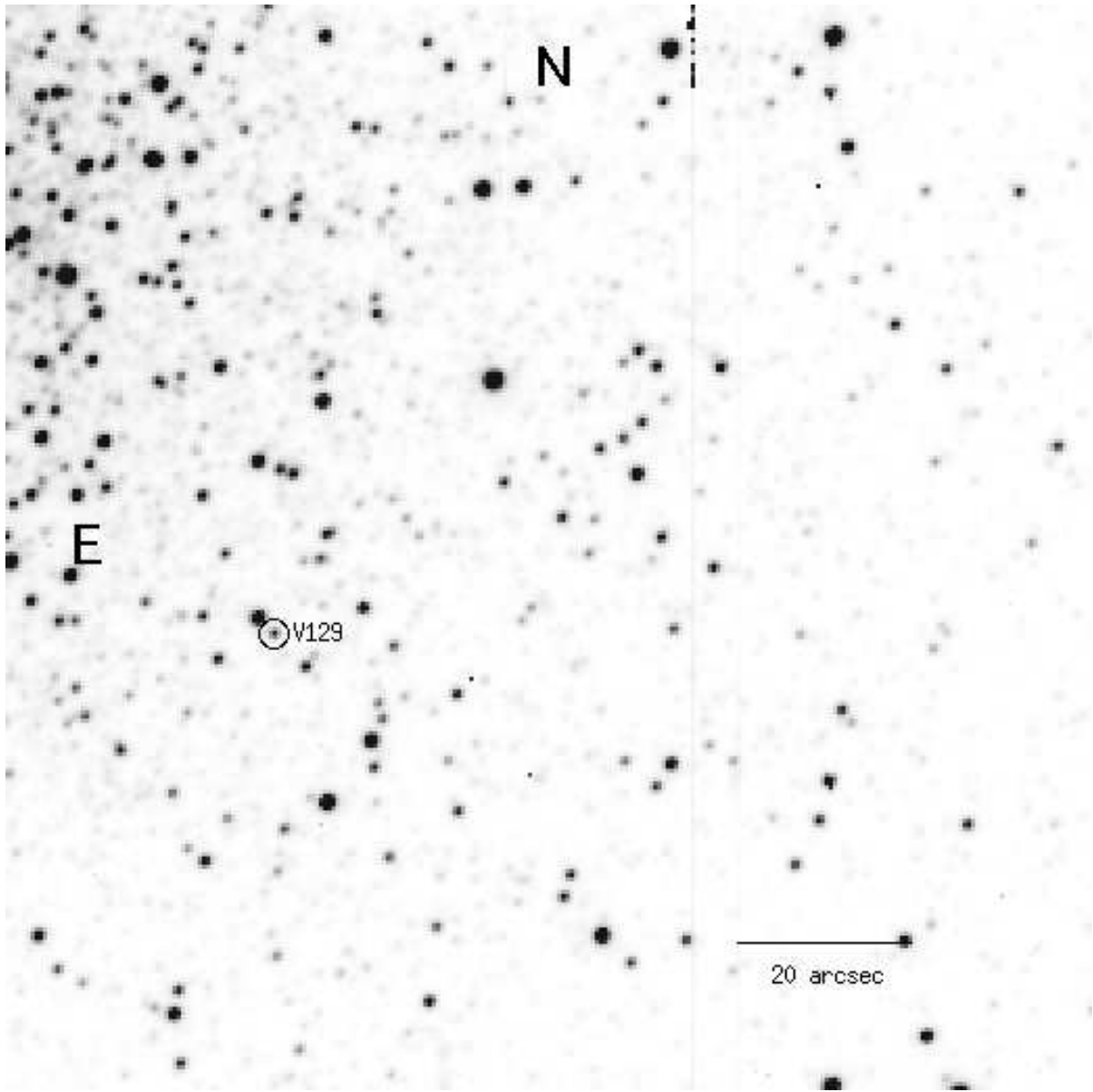
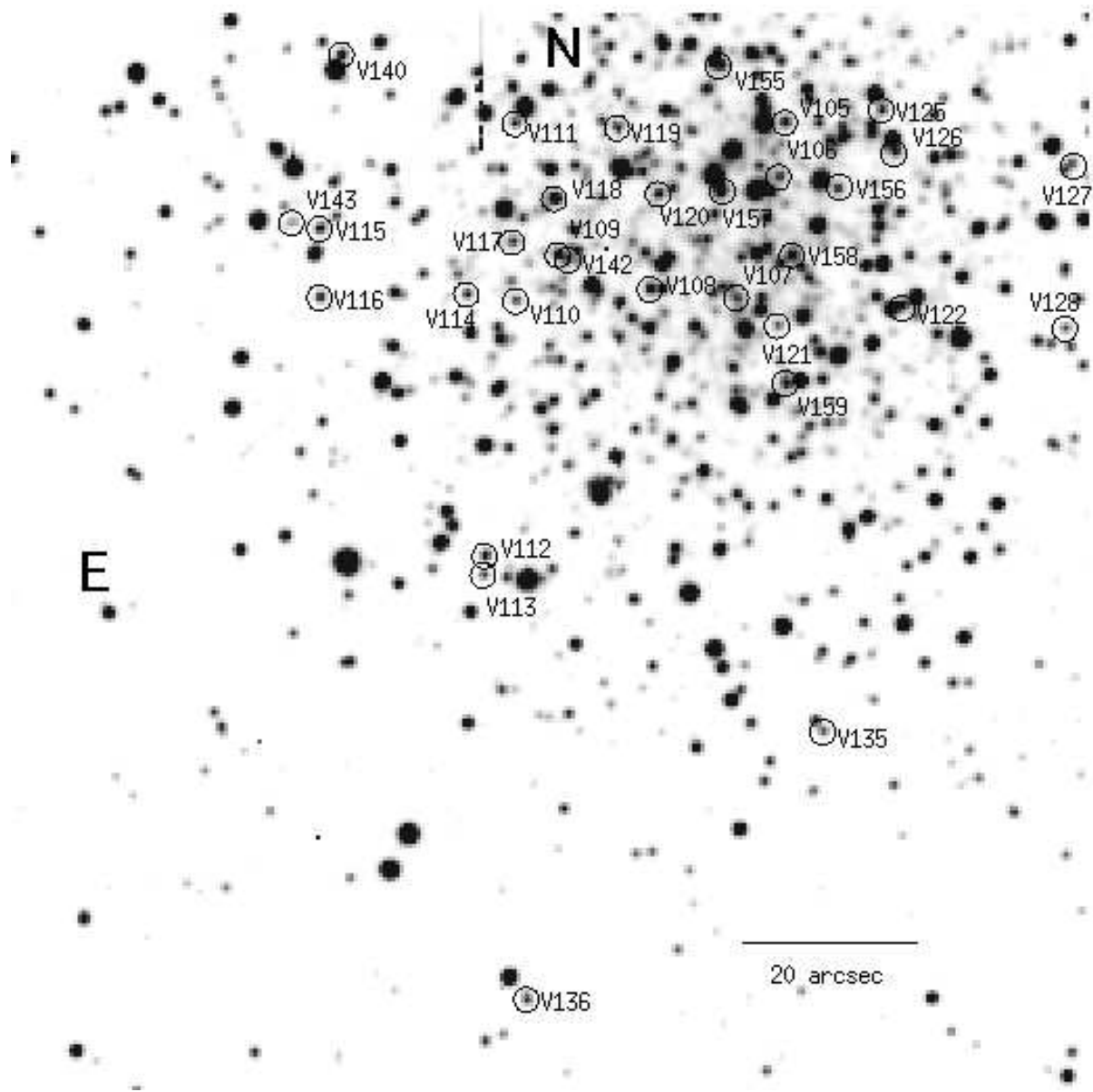
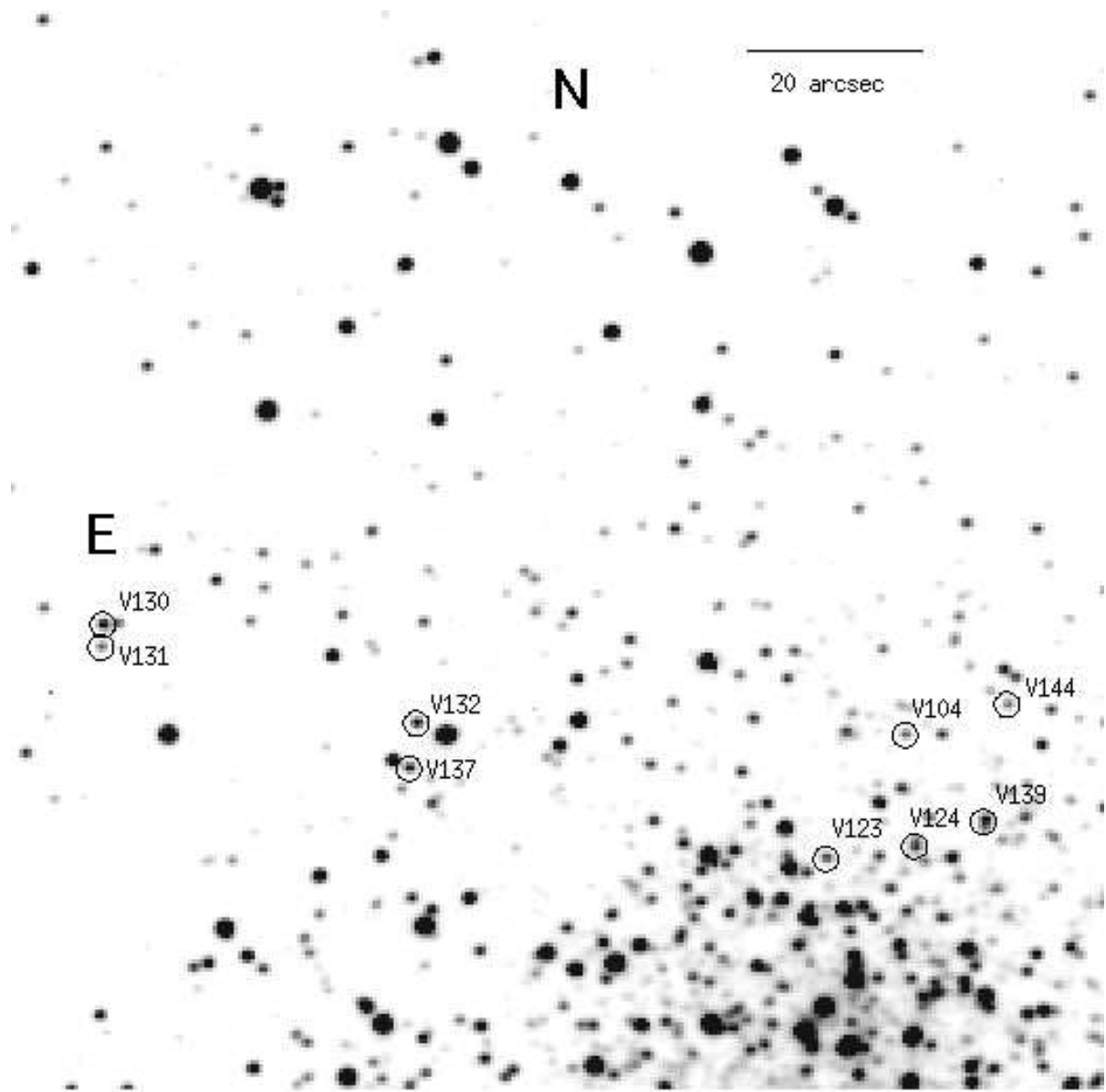
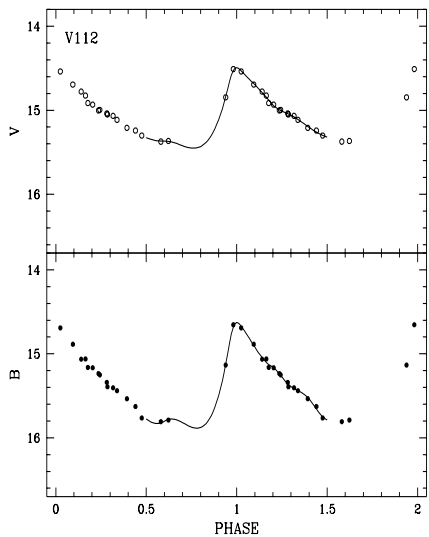
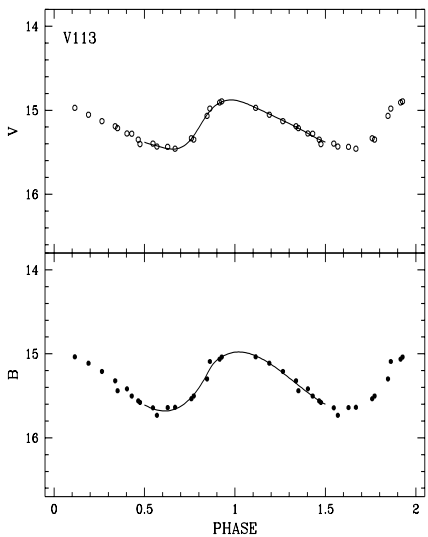
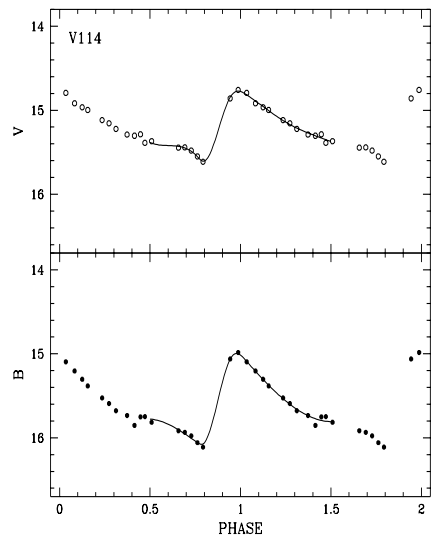
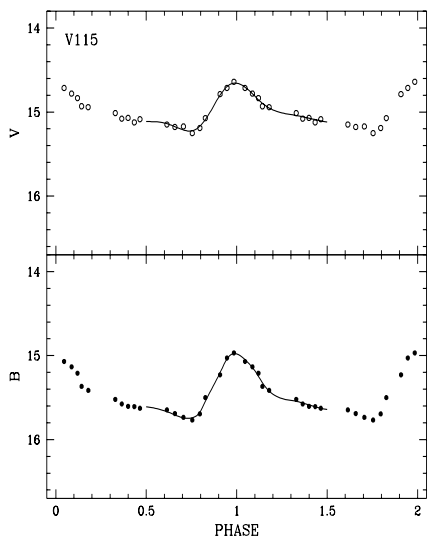
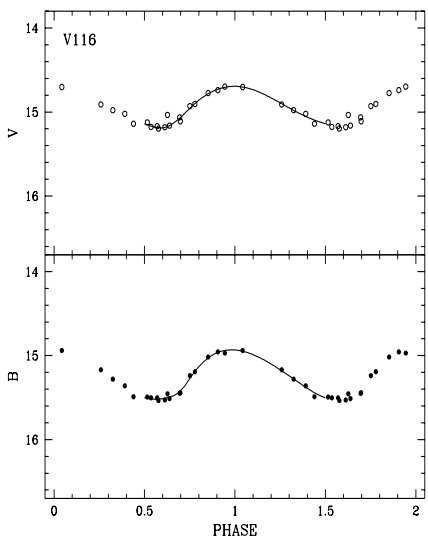
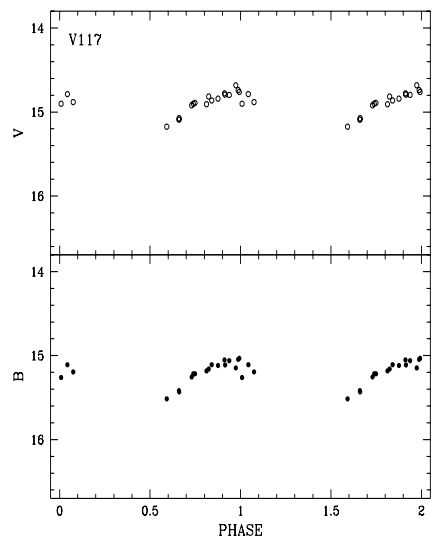
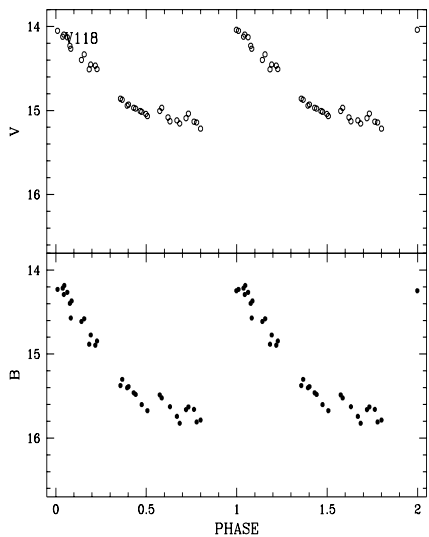
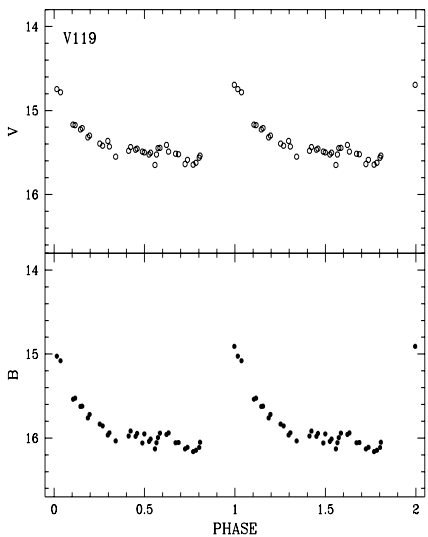
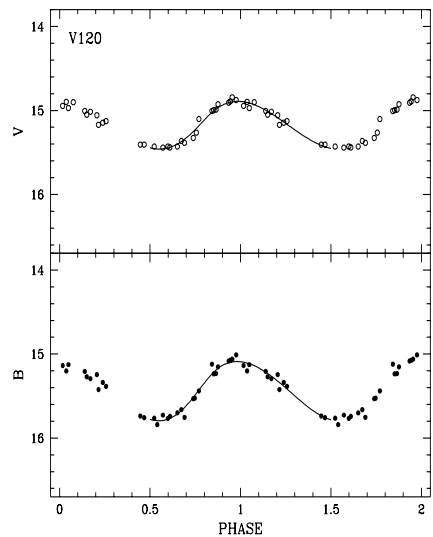
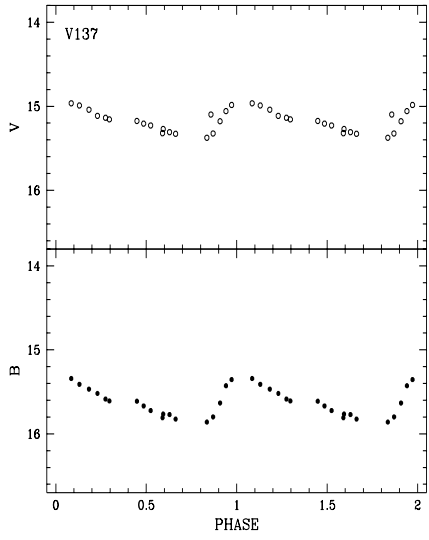
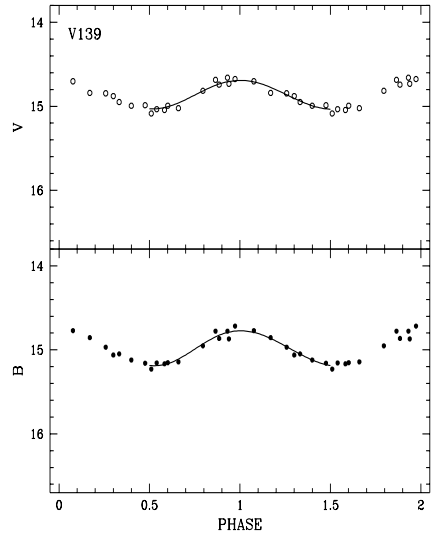
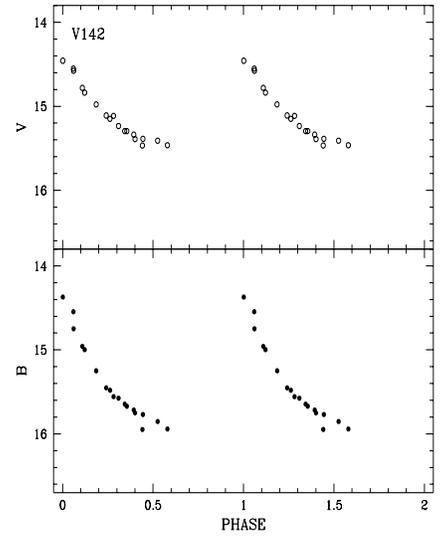
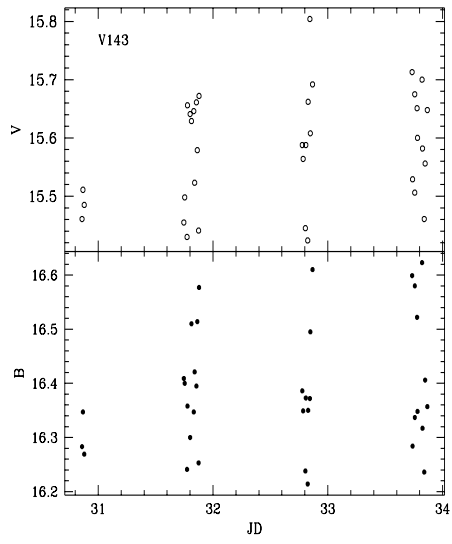
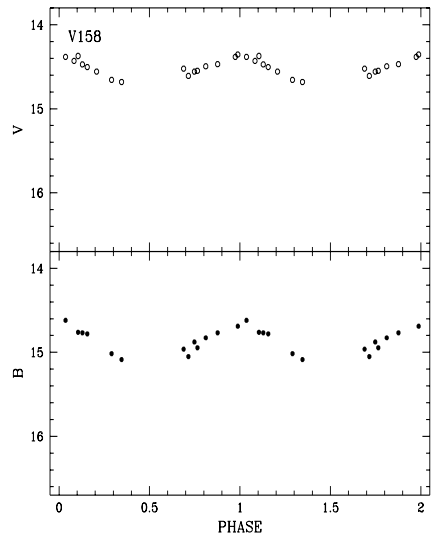
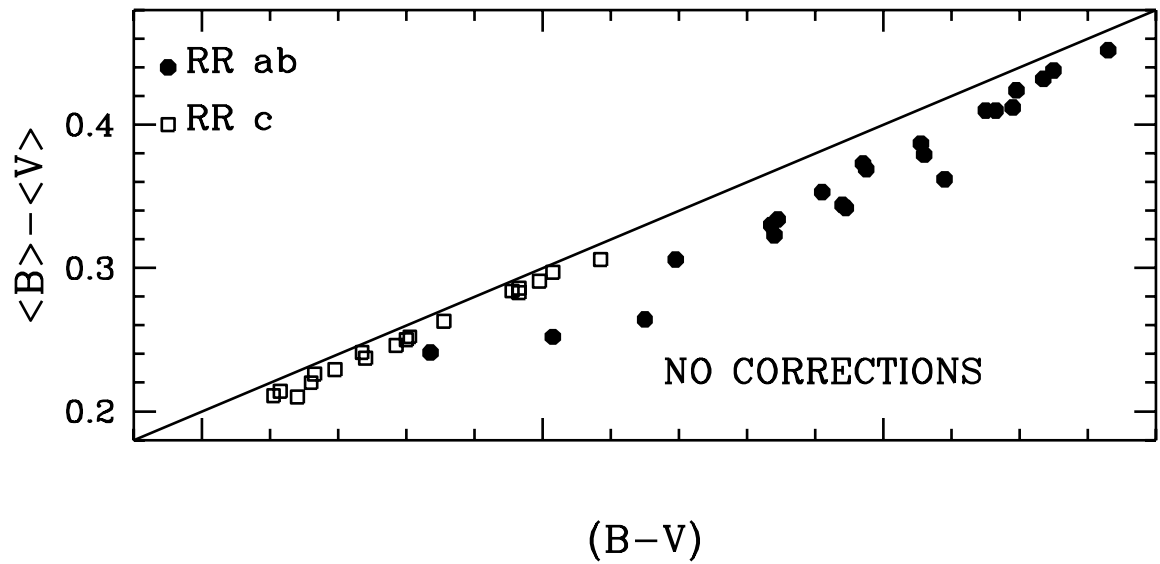
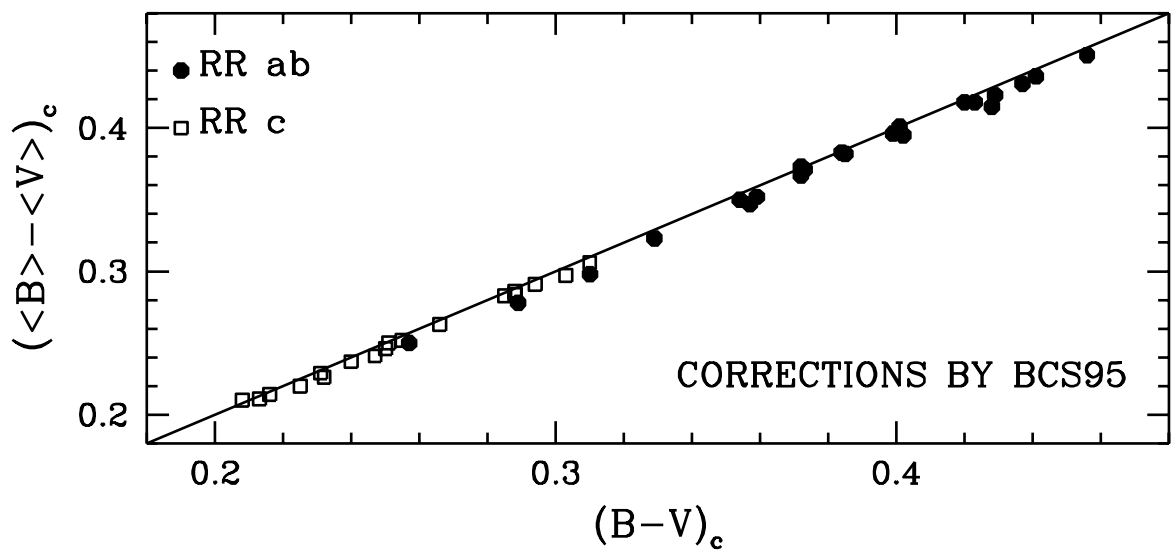
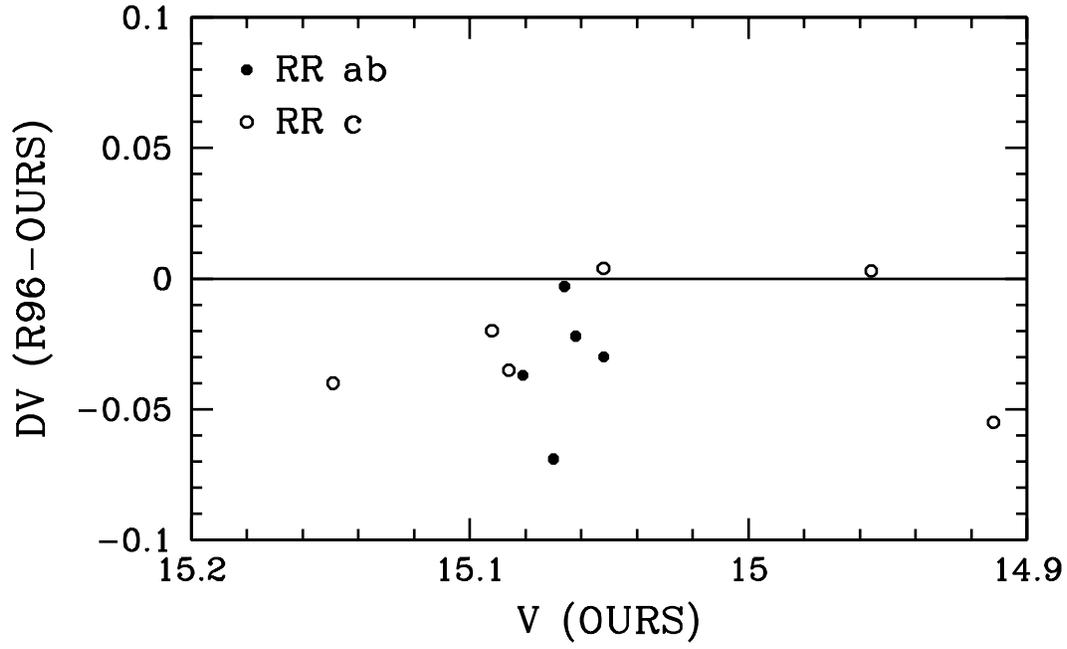
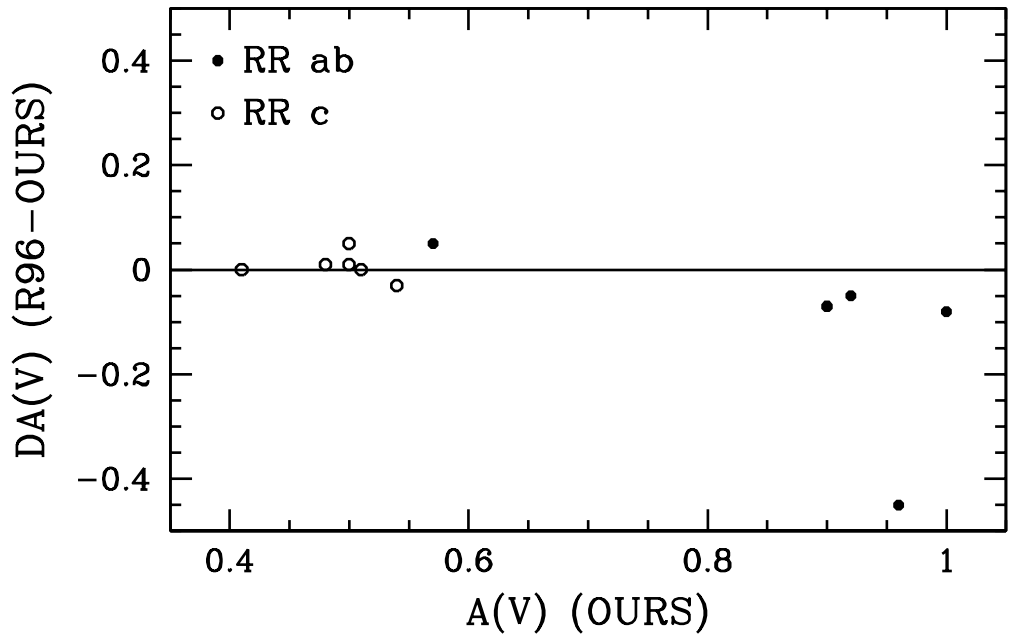
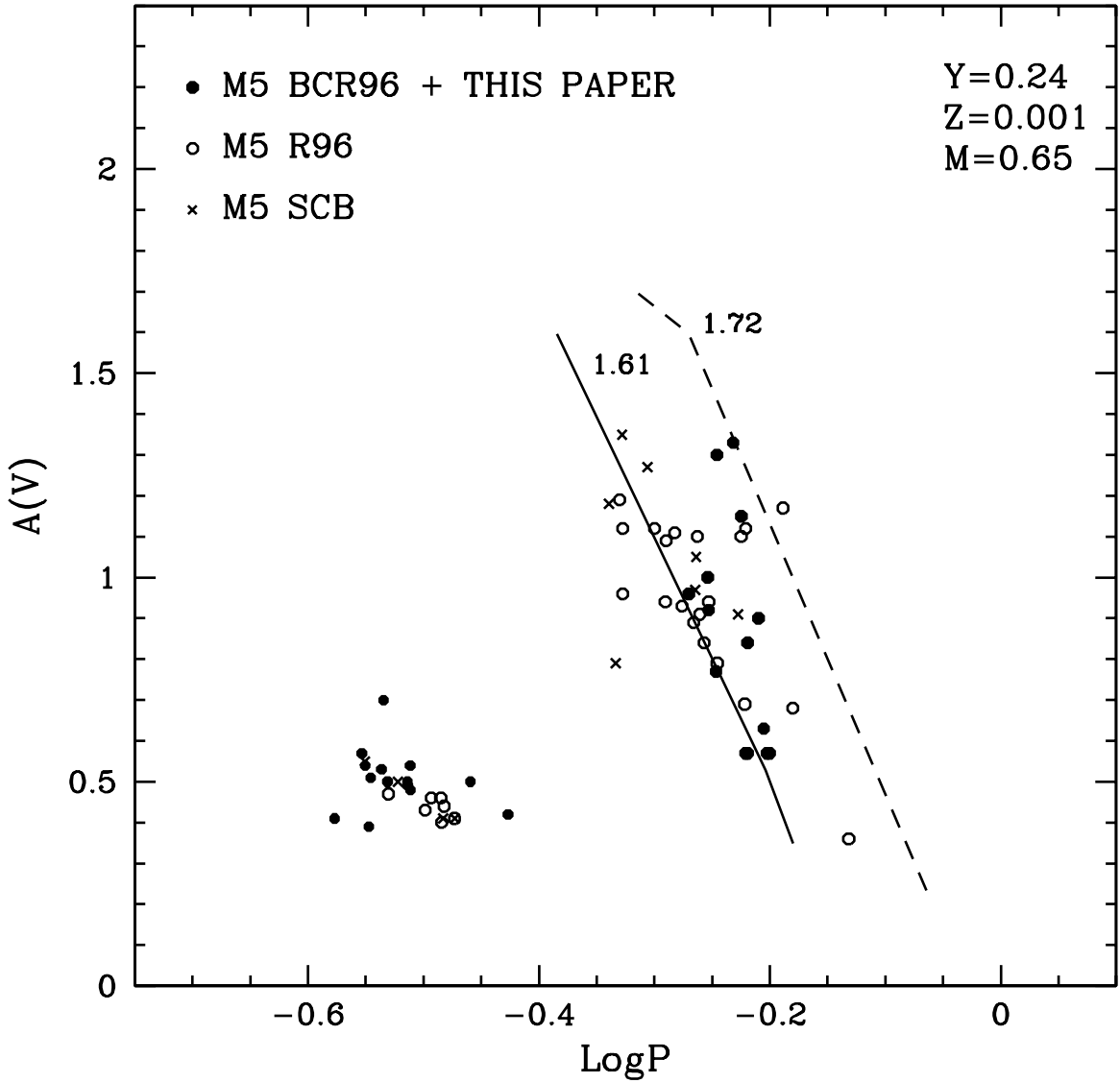
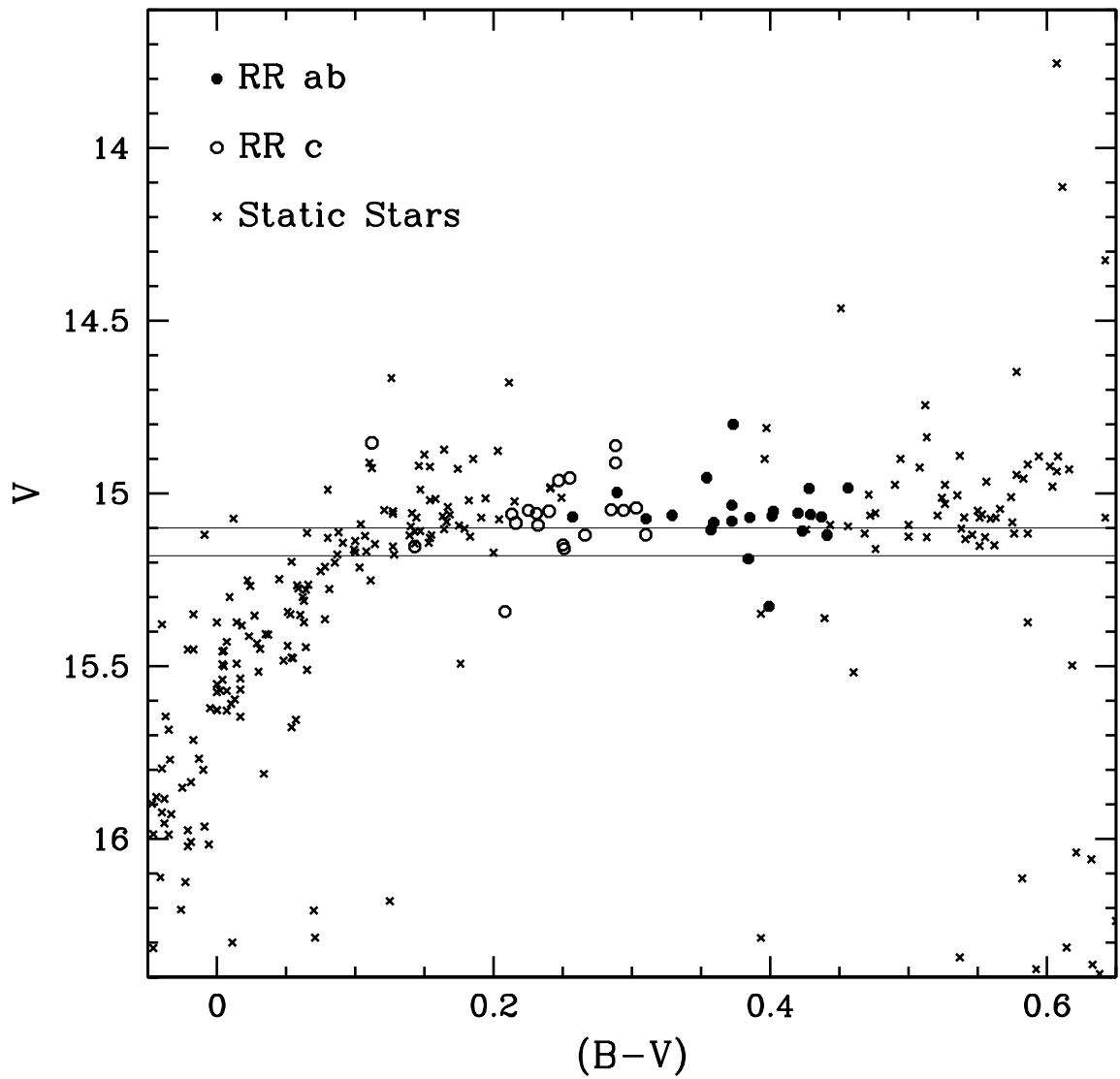
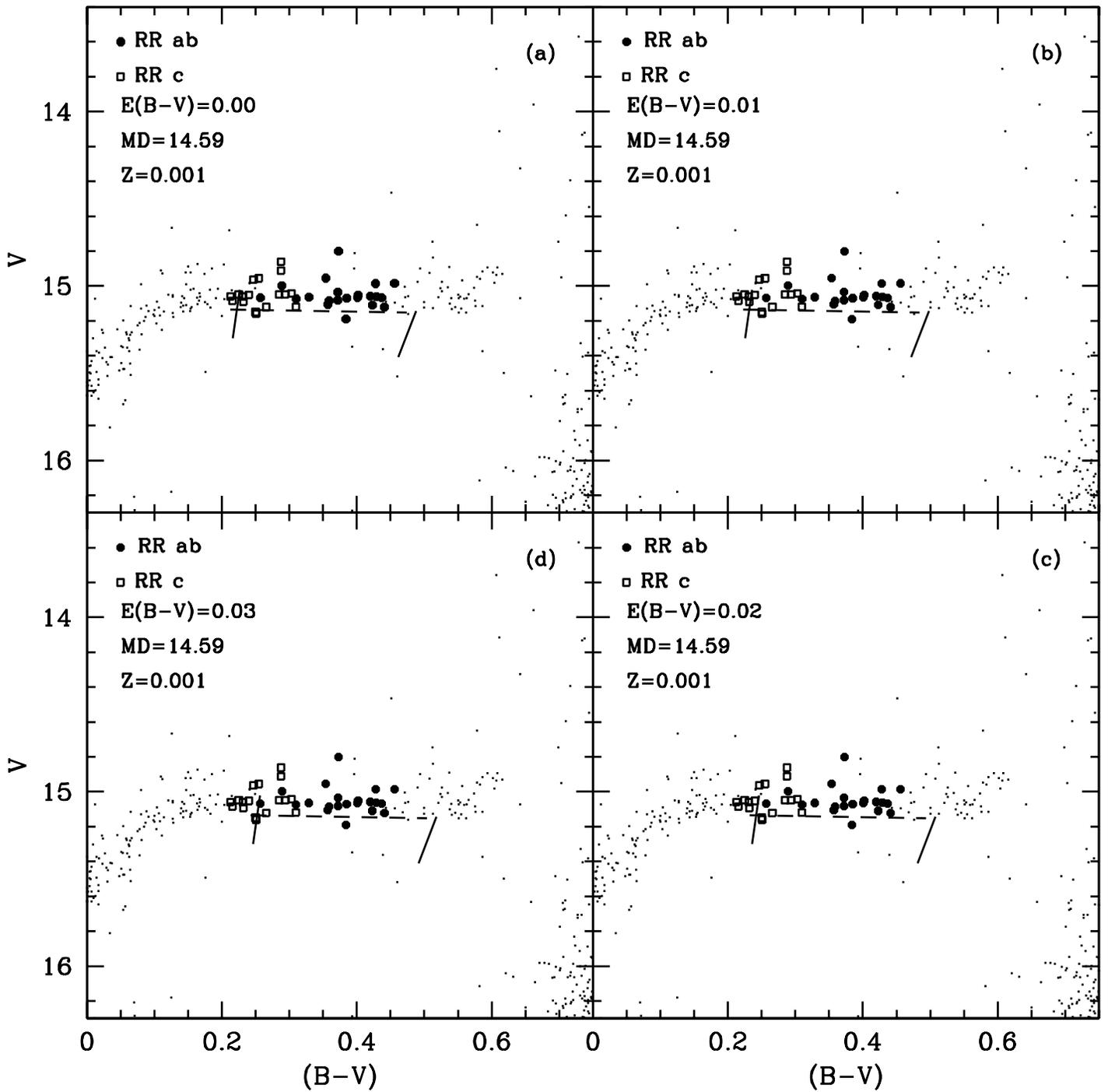
\end{document}